\documentclass[prd,twocolumn,superscriptaddress,nofootinbib,amsmath,amssymb]{revtex4}
\usepackage{graphicx}
\usepackage{dcolumn}
\usepackage{lscape}
\usepackage{bm}
\usepackage{latexsym,epsfig}
\usepackage[usenames,dvipsnames]{color}
\usepackage{hyperref} 
\usepackage{verbatim}  
\usepackage{soul}   
 \usepackage{graphicx}
 \usepackage[font=footnotesize]{subfig} 
 \usepackage{relsize}
 \usepackage{subfig}

\def\bigD{{\cal D}}

\def\beq{\begin{equation}}
\def\eeq{\end{equation}}
\def\bea{\begin{eqnarray}}
\def\eea{\end{eqnarray}}

\def\eqref#1{Eq.~(\ref{eq:#1})}
\newcommand*{\eqlab}[1]{\label{eq:#1}}
\newcommand*{\figref}[1]{Fig.~\ref{fig:#1}}
\newcommand*{\figlab}[1]{\label{fig:#1}}

\newcommand*{\secref}[1]{Section~\ref{sec:#1}}
\newcommand*{\seclab}[1]{\label{sec:#1}}

\begin{document}


\title{Coherent radio emission from the electron beam sudden appearance}

\author{Krijn D. de Vries\footnote{krijndevries@gmail.com}}
\affiliation{Vrije Universiteit Brussel, Dienst ELEM, IIHE, Pleinlaan 2, 1050, Brussel, Belgium}

\author{Michael DuVernois}
\affiliation{Dept. of Physics and Wisconsin IceCube Particle Astrophysics Center, University of
Wisconsin, Madison, WI 53706, USA}
\author{Masaki Fukushima}
\affiliation{ICRR, University of Tokyo, 5-1-5 Kashiwanoha, Kashiwa, Chiba 277-8522, Japan}
\author{Romain Ga\"{i}or\footnote{romain.gaior@lpnhe.in2p3.fr}}
\affiliation{Department of Physics, Chiba University, Yayoi-cho 1-33, Inage-ku, Chiba 263-8522, Japan}
\affiliation{Laboratoire de Physique Nucl\'eaire et de Hautes Energies (LPNHE), Sorbonne Universit\'e, Universit\'e Paris Diderot, CNRS-IN2P3, France}
\author{Kael Hanson}
\affiliation{Dept. of Physics and Wisconsin IceCube Particle Astrophysics Center, University of
Wisconsin, Madison, WI 53706, USA}
\author{Daisuke Ikeda\footnote{ikeda@icrr.u-tokyo.ac.jp}}
\affiliation{ICRR, University of Tokyo, 5-1-5 Kashiwanoha, Kashiwa, Chiba 277-8522, Japan}
\affiliation{Earthquake Research Institute, University of Tokyo, 1-1-1 Yayoi,
Bunkyo-ku, Tokyo 113-0032, Japan}
\author{Yusuke Inome}
\affiliation{Faculty of Science and Engineering, Konan University, Kobe 658-8501, Japan}
\affiliation{ICRR, University of Tokyo, 5-1-5 Kashiwanoha, Kashiwa, Chiba 277-8522, Japan}
\author{Aya Ishihara}
\affiliation{Department of Physics, Chiba University, Yayoi-cho 1-33, Inage-ku, Chiba 263-8522, Japan}
\author{Takao Kuwabara}
\affiliation{Department of Physics, Chiba University, Yayoi-cho 1-33, Inage-ku, Chiba 263-8522, Japan}
\author{Keiichi Mase\footnote{mase@hepburn.s.chiba-u.ac.jp}}
\affiliation{Department of Physics, Chiba University, Yayoi-cho 1-33, Inage-ku, Chiba 263-8522, Japan}
\author{John N. Matthews}
\affiliation{University of Utah,  Department of Physics and Astronomy, High Energy Astrophysics Institute, Salt Lake City, UT 84112, USA}
\author{Thomas Meures}
\affiliation{Dept. of Physics and Wisconsin IceCube Particle Astrophysics Center, University of
Wisconsin, Madison, WI 53706, USA}
\author{Pavel Motloch}
\affiliation{Kavli Institute for Cosmological Physics \& Department of Physics, University of Chicago, Chicago, IL 60637, USA}
\author{Izumi S. Ohta}
\affiliation{Faculty of Science and Engineering, Konan University, Kobe 658-8501, Japan}
\author{Aongus O'Murchadha}
\affiliation{Dept. of Physics and Wisconsin IceCube Particle Astrophysics Center, University of
Wisconsin, Madison, WI 53706, USA}
\author{Florian Partous}
\affiliation{Vrije Universiteit Brussel, Dienst ELEM, IIHE, Pleinlaan 2, 1050, Brussel, Belgium}
\author{Matthew Relich}
\affiliation{Department of Physics, Chiba University, Yayoi-cho 1-33, Inage-ku, Chiba 263-8522, Japan}
\author{Hiroyuki Sagawa}
\affiliation{ICRR, University of Tokyo, 5-1-5 Kashiwanoha, Kashiwa, Chiba 277-8522, Japan}
\author{Tatsunobu Shibata}
\affiliation{High Energy Accelerator Research Organization (KEK), 2-4 Shirakata-Shirane, Tokai-mura,
Naka-gun, Ibaraki, 319-1195, Japan}
\author{Bokkyun Shin}
\affiliation{Osaka City University, Osaka 558-8585, Japan}
\author{Gordon Thomson}
\affiliation{University of Utah,  Department of Physics and Astronomy, High Energy Astrophysics Institute, Salt Lake City, UT 84112, USA}
\author{Shunsuke Ueyama}
\affiliation{Department of Physics, Chiba University, Yayoi-cho 1-33, Inage-ku, Chiba 263-8522, Japan}
\author{Nick van Eijndhoven}
\affiliation{Vrije Universiteit Brussel, Dienst ELEM, IIHE, Pleinlaan 2, 1050, Brussel, Belgium}
\author{Tokonatsu Yamamoto\footnote{yamamoto.tokonatu@gmail.com}}
\affiliation{Faculty of Science and Engineering, Konan University, Kobe 658-8501, Japan}
\affiliation{ICRR, University of Tokyo, 5-1-5 Kashiwanoha, Kashiwa, Chiba 277-8522, Japan}
\author{Shigeru Yoshida}
\affiliation{Department of Physics, Chiba University, Yayoi-cho 1-33, Inage-ku, Chiba 263-8522, Japan}

\begin{abstract}
We report on the measurement of coherent radio emission from the electron beam sudden appearance at the Telescope Array Electron Light Source facility. This emission was detected by four independent radio detector setups sensitive to frequencies ranging from 50~MHz up to 12.5~GHz. We show that this phenomenon can be understood as a special case of coherent transition radiation by comparing the observed results with simulations. The in-nature application of this signal is given by the emission of cosmic ray or neutrino induced particle cascades traversing different media such as air, rock and ice.\\\\
\end{abstract}

\maketitle

\section{Introduction}
The flux of cosmic particles decreases dramatically with energy and even 
the currently largest cosmic ray detectors, the Pierre Auger 
Observatory~\cite{ThePierreAuger:2015rma}, and Telescope Array 
detector~\cite{tafd,tasd}, run low in statistics above a few $10^{19}$~eV, 
while the IceCube neutrino observatory~\cite{Aartsen:2016nxy} loses 
sensitivity above a few PeV. 
\\The use of the emission of cosmic-ray or neutrino induced particle cascades at radio frequencies has been studied for several decades and could provide a cost effective technique to increase the exposure thanks to the long attenuation length of radio waves. High-energy cosmic-ray and neutrino induced particle cascades are known to emit radio waves through the time variation of a net charge excess in the cascade (the Askaryan effect) and the transverse current induced by the geomagnetic field~\cite{Askaryan,Kahn,Olaf}. The combination of these emission mechanisms with Cherenkov effects~\cite{cher1,cher2} leads to coherent radio emission in the MHz-GHz frequency range. Those mechanisms are well studied and are the base of several \textit{insitu} experiments probing the cosmic particle flux (see ~\cite{radioreview1,radioreview2} for a review). \\
Searches for other exploitable radio emission mechanisms and detection methods are underway. The ELS (Electron Light Source)~\cite{ELS} is a source of relativistic electrons, located in the desert of Utah on the site of the Telescope Array (TA) detector. Electron bunches are shot up in the air and used to mimic an air shower for the calibration of the TA fluorescence detector. It is also an ideal setup to test the radio emissions from particle cascades. Therefore, in the last few years, several radio detection systems were set up to explore techniques such as the radar detection method, namely AirRadar~\cite{AirRadar} and IceRadar~\cite{IceRadar}. Also direct radio emission mechanisms were probed, namely the Molecular Bremsstrahlung emission by the Konan experiment~\cite{Konan}, and the Askaryan emission in ice with the ARAcalTA experiment~\cite{ARAcalTA}. 
\\Besides the signal sought for, all four experiments have observed a clear radio signal as the electron beam exits the metallic container, inside which the beam pipe is located. The detected signal is not explained by the Askaryan nor the geomagnetic effect, but is related to sudden change of the electromagnetic potential as observed by the radio detection setups. In this work, we show that this effect can be seen as a special case of transition radiation where one medium, the inside of the container, is completely opaque, and the other medium, the air, is transparent. 
\\A similar signal was first observed during a test beam experiment~\cite{Argonne} in the year 2000 when the electron beam crossed an aluminum-air boundary at the beam exit. Another measurement was done following this experiment, which is described in~\cite{SLACsalt}. However, the obtained coherent transition radiation signal was never completely quantified. \\In this paper we give a detailed characterization of the signal observed from the electron beam sudden appearance, based on the data collected by four experiments set at the TA-ELS facility. These experiments were operating over a wide frequency range between 50 MHz and 12.5 GHz. We show that the sudden appearance signal is a special case of coherent transition radiation. As such, the agreement of the obtained data with simulation provides confidence in the understanding of the coherent transition radiation as predicted for the in-nature process of a high-energy cosmic-ray or neutrino induced particle cascade traversing different media. Indeed, as shown recently, a detectable radio signal is expected as an air shower is absorbed by the ground~\cite{extasis} and for neutrino induced cascades traversing the ground to air boundary~\cite{CosmicRayRadio,Motloch}.
\\After reviewing the mechanism of coherent transition radiation in~\secref{sec:theory} we describe the four concerned experiments and the test beam setup in~\secref{experiments}. In ~\secref{data-sim} the sudden appearance signal is characterized over the full frequency range. The simulation methods are also presented and compared with the observations. We conclude on the application and the scope of the presented results.

\section{From sudden appearance to coherent transition radiation}
\seclab{sec:theory}
To understand the electron beam sudden appearance emission described in this work, we will consider the Li\'enard-Wiechert potentials from classical electrodynamics,
\beq
A^{\mu}(\vec{x},t)=\left.\frac{J^{\mu}(t')}{{\cal D}}\right|_{ret}.
\eeq
The potential observed at location $\vec{x}$ and observer time $t$ has to be evaluated at the emission time $t'$. The emission source is given by the four current $J^{\mu}$, which contains the net charge and currents that are considered with $\mu=0$ indicating the scalar potential and $\mu=1-3$ denoting the current in the $\vec{x}$ direction. The geometry of the process is contained in the retarded distance which for a homogeneous medium is given by ${\cal D}=R(1-n\beta\cos(\theta))$. Here $R$ denotes the distance from the emission point to the observer, the refractive index of the medium is given by $n$, the speed of the particle cascade with respect the speed of light is given by $\beta=v/c$, and $\theta$ gives the opening angle between the axis of propagation and the line of sight towards the observer. 

Since we discuss the situation of a particle cascade traversing different media, we have to consider the more general expression for the retarded distance given by,
\beq
{\cal D}=L\left|\frac{dt}{dt'}\right|.
\eeq
Here $L$ denotes the optical path length from the charge to the observer. To understand transition radiation, we have to investigate the derivative $dt'/dt$, given by the relation between the emission time as function of the observer time $t'(t)$. 
\\Following~\cite{CosmicRayRadio}, an example of this relation is given in~\figref{e-field}. Here we consider a perpendicular incoming, $10^{17}$~eV cosmic-ray air shower hitting an air-ice boundary at 3~km above sea level, $n_{air}=1.0003$; $n_{ice}=1.3$. The observer is located 100~m deep in the ice, at a horizontal distance of 240~m from the shower axis. Defining $t=t'=0$ as the time when the shower crosses the observer plane perpendicular to the shower axis, in~\figref{e-field} we show the relation between the emission height $z=-ct'$, which scales linearly with the emission time, $t'$, and the observer time $t$. 
\begin{figure}[!ht]
  \centering
  \hspace*{-3ex}
  \includegraphics[width=.9\linewidth]{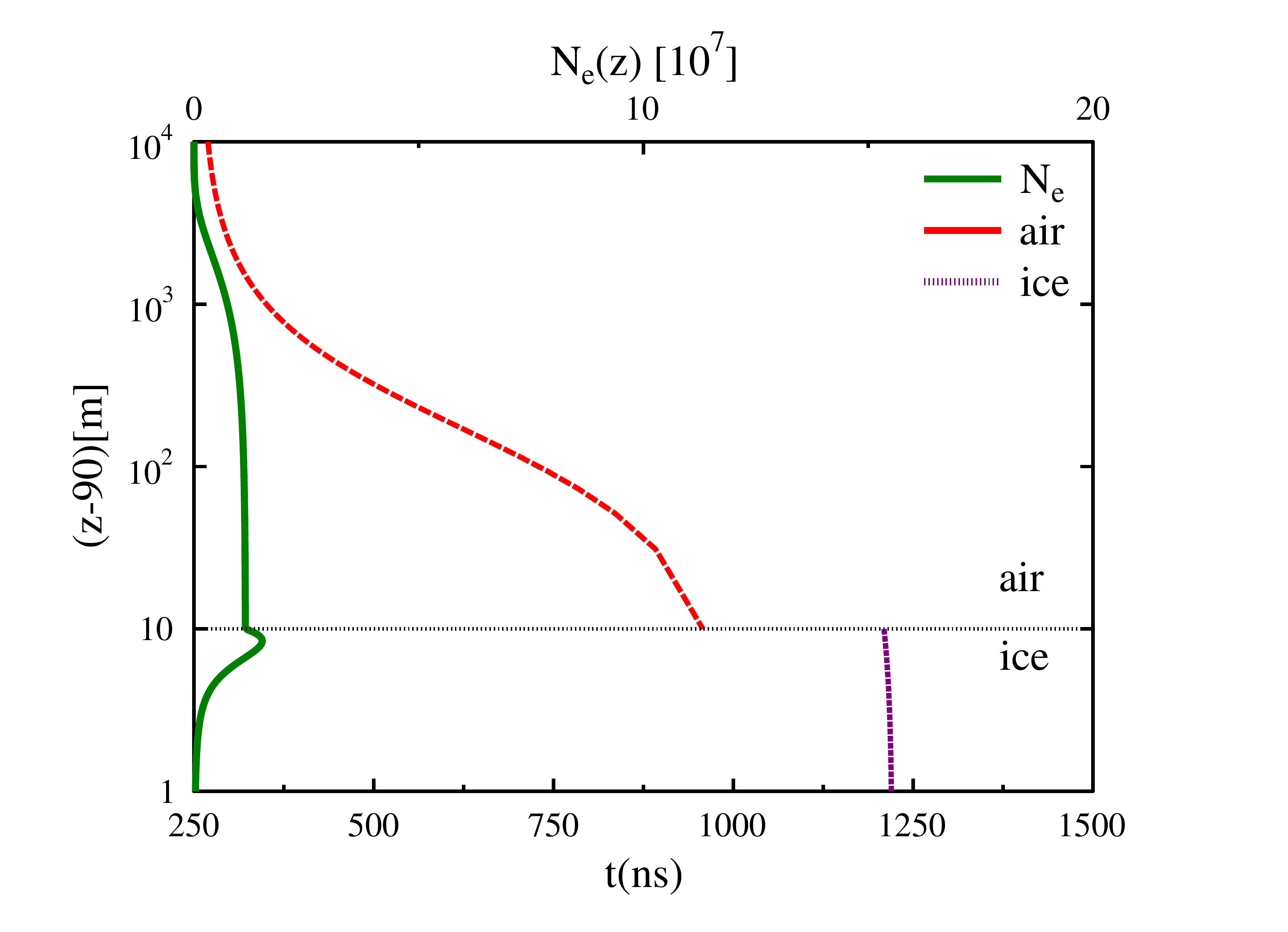}
  \caption{The emission height, plotted as function of the observer time for a $10^{17}$~eV, perpendicular incoming cosmic-ray air shower hitting an ice surface at 3~km height above sea level. The observer is located 100~m below the air-ice boundary, at a horizontal distance of 240~m from the shower axis. The dashed red line denotes the emission in air, the striped purple line gives in-ice emission. The total number of particles denoted on the top-axis is given by the full green line. Figure taken from~\cite{CosmicRayRadio}.}
  \figlab{e-field}
\end{figure}
The (dashed) red line indicates the emission height, scaling directly to the emission time, as function of the observer time in air. The (striped) purple line indicates the emission height as function of the observer time for the in-ice emission. 
\\It should be noted that, following Eqs. 1) and 2), the electric field scales directly to the derivative of these lines given by $dz/dt \propto dt'/dt$, which is the so-called compression, or beaming factor. This immediately gives an intuitive picture of the Cherenkov effect, which occurs when the derivative $dz/dt$ diverges. This for example occurs for the (Askaryan) emission in ice given by the purple line in~\figref{e-field}. In this situation, a finite part of the cascade is observed by the observer at once, and hence the observed field is amplified greatly.
\\Transition radiation considers an even more extreme geometrical situation, where at the boundary the derivative $dz/dt \propto dt'/dt$ is discontinuous. It follows that equivalent to the Cherenkov effect, transition radiation is a purely geometrical effect. To illustrate this, in~\figref{e-field} we consider a very interesting situation in which the signal emitted just above the boundary arrives at a completely different time with respect to the signal emitted directly below the boundary due to their different travel paths. It follows that at the boundary, the retarded distance $|\bigD|_{z=z_b}\propto dz_b/dt$, and the corresponding electric field is ill defined (shown by the discontinuity between the (dashed) red and (striped) purple lines in~\figref{e-field}). This immediately leads to the question: How can we properly describe transition radiation? Following the approach given in~\cite{CosmicRayRadio}, taking the limit to the boundary from both sides, for a single electron, one obtains,
\beq
E_{tr}^i(t,\vec{x})=\frac{e\delta(z-z_b)}{4\pi\epsilon_0 c}\lim_{\epsilon\rightarrow 0}\left( \frac{x^i}{|\bigD|^2_{z_b+\epsilon}} -\frac{x^i}{|\bigD|^2_{z_b-\epsilon}} \right).
\eqlab{etrr}
\eeq
It immediately follows that if $n_1=n_2$, and hence no boundary is crossed, $|\bigD|^2_{z_b-\epsilon}=\bigD|^2_{z_b+\epsilon}$, the field vanishes as it should. Following this approach, transition radiation is obtained from the interference between the sudden death signal of the emission vanishing in air directly above boundary at $z_b^+=z_b+\epsilon$, given by the term scaling to $1/|\bigD|^2_{z_b+\epsilon}$, and the sudden appearance signal in ice observed directly below the boundary at $z_b^-=z_b-\epsilon$, given by the term scaling to $1/|\bigD|^2_{z_b-\epsilon}$. More interestingly, for the situation considered in~\figref{e-field}, the sudden death of the cascade is observed at a different time as the sudden appearance. 
\begin{figure}[!ht]
  \centering
  \hspace*{-3ex}
  \includegraphics[width=.9\linewidth]{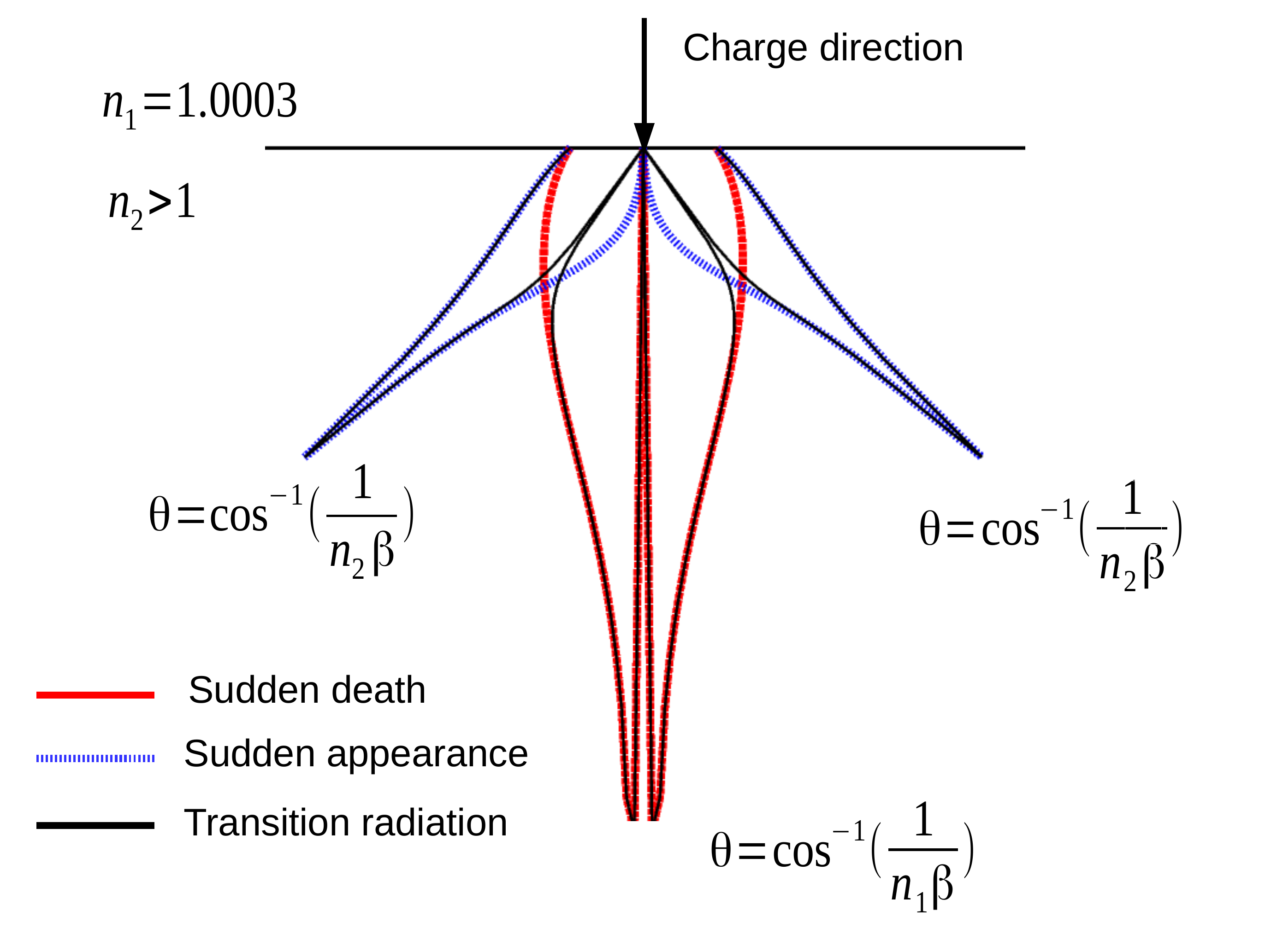}
  \caption{The angular distribution showing the logarithm of the electric field strength as function of emission angle in the forward direction for a charge crossing from air into a dense medium. The transition radiation (full black line) is composed out of the interference between the sudden death emission in air (dashed red line) and the sudden appearance in ice (striped blue line). A quantitative calculation of the emission strength is presented in Figs.~2 and 6 of~\cite{CosmicRayRadio}.}
  \figlab{angular}
\end{figure}

The angular dependence of the transition radiation is illustrated in~\figref{angular} for a charge crossing from air into a dense medium (full black line). In this figure, we also show the sudden death (dashed red line) and sudden appearance (striped blue line) components separately, where the transition radiation is given by the interference between both components. The angular distribution of the transition radiation signal is understood from~\eqref{etrr}. Both the sudden death signal in air as well as the sudden appearance signal in ice peak close to the Cherenkov angle in their corresponding medium. For small observer angles with respect to the cascade axis, the sudden death and sudden appearance signal show destructive interference, for large angles close to the horizon, due to the time separation between both signals, no interference occurs.
\subsubsection{The electron beam sudden appearance}
The difference between conventional transition radiation and the sudden appearance emission considered in this work is illustrated in~\figref{SAscheme2}. For the situation described in this work, the sudden death component vanishes, since the electron beam will be shielded by the metallic walls of the accelerator housing. Here 'shielding' means that the refractive index of the metallic walls is completely imaginary at radio wavelengths. Hence, the waves get partially reflected and partially absorbed and will not be able to reach the observer for the considered geometries. More formally, the potentials, directly scaling to $A\propto dt'/dt$, vanish while the beam propagates inside the accelerator.
\begin{figure}[!th]
  \centering
  \hspace*{-3ex}
  \includegraphics[width=0.9\linewidth]{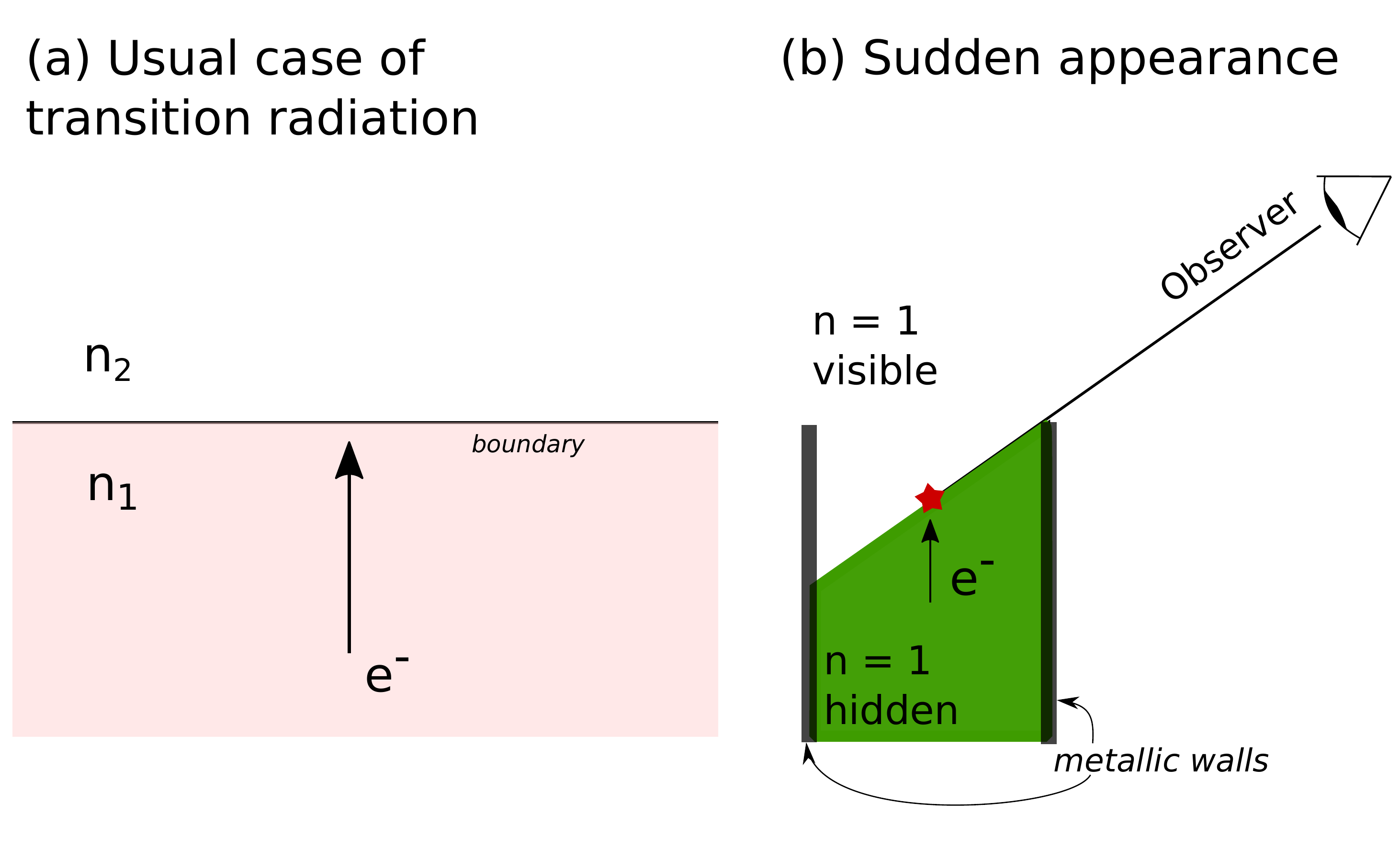}
  \caption{a) The situation conventionally considered for transition radiation, given by a charge moving between different media with refractive indices $n_1$, and $n_2$, crossing a boundary surface. b) The situation considered in this work. Even though the electron beam does not cross a boundary surface, it moves from the accelerator container, where it is effectively hidden for the observer, to free space.}
  \figlab{SAscheme2}
\end{figure}
It follows that even though both situations outlined in~\figref{SAscheme2} seem very different, we observe a special situation where the sudden death signal vanishes and only the electron beam sudden appearance signal is observed. 

\section{Beam experiments at the Telescope Array Electron Light Source facility}
\seclab{experiments}
The Telescope Array (TA) experiment is the largest ultra-high energy cosmic ray observatory in the northern hemisphere. Cosmic-ray air showers are observed by detecting either the particle footprint when the cascade hits Earth's surface, or by detecting the fluorescence light emitted from exited air molecules. For the absolute energy calibration of the fluorescence detector, a linear electron accelerator, the Telescope Array Electron Light Source (TA-ELS), was installed at a distance of 100~m from one of the fluorescence detector stations. The TA-ELS accelerates electrons to a mean energy of 40~MeV. The electrons leave the beam pipe at a distance of 122.5~cm below the container exit point (See Fig.~\ref{fig:scheme}). The beam is directed upward into the atmosphere to mimic an electromagnetic shower, with a repetition frequency of 0.5~Hz. The electrons reach a height of roughly 130 m. The output charge, given by the total number of electrons in each beam shot, is variable with a maximum charge of $\sim 10^9\;e^-$, and is monitored by a Wall Current Monitor (WCM) installed on the wall of the beam pipe exit.

To extend the frequency range up to which coherent radio emission is observed, the beam length has been altered from the TA-ELS default value of 1~$\mathrm{\mu s}$ to either 1.5~ns or 5~ns full width half maximum for the experiments described in this work. The accelerator tube used for the TA-ELS is optimized for the S-band (2.856~GHz). It follows that the beam is composed of multiple electron sub-bunches with a 350~ps interval, and the lateral beam spread is found to be of the order of a few centimeters. In order to obtain the beam bunch structure, a Faraday Cup (FC) was used~\cite{ELSSystematic}. The absolute beam charge for each beam is calibrated with an accuracy of 3.3~\%, by taking the correlation between the charge measured by the FC and the WCM.

\subsubsection{Radio detection setups}

\begin{figure}[!th]
  \centering
  \hspace*{-3ex}
  \includegraphics[width=0.85\linewidth]{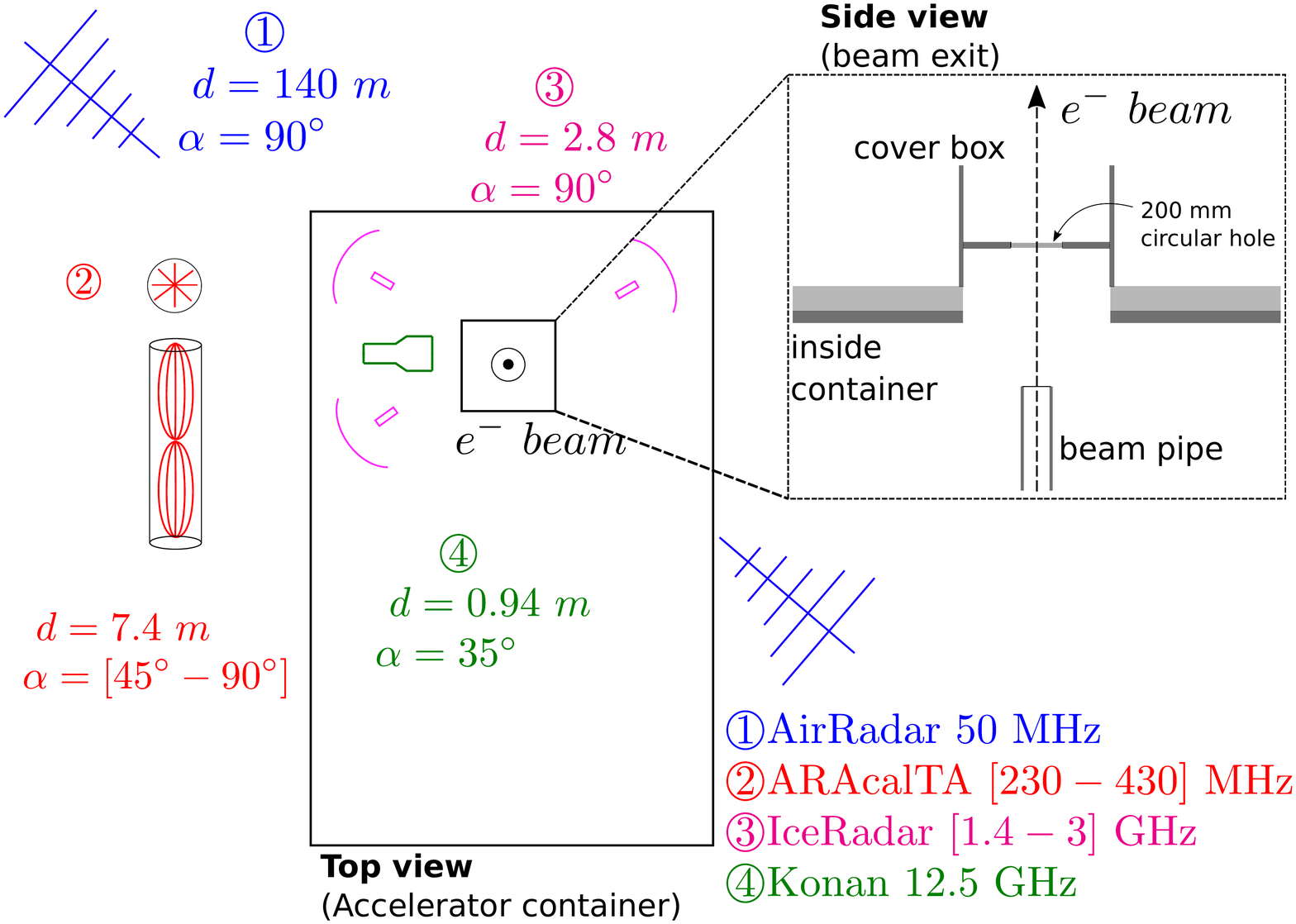}
  \caption{The TA-ELS facility and the different radio experiments located at a distance $d$ from the beam exit point is shown in the top view. The observation angle $\alpha$ with respect to the beam direction is also shown for each experiment. The details of the beam exit configuration are presented in the side view.}
  \label{fig:scheme}
\end{figure}
The radio signal from the electron beam sudden appearance is observed by four different experiments at the TA-ELS site. An overview of the different experimental setups is given in~\figref{scheme}. The antennas were located at a distance $d$ from the beam exit point. The observer angle $\alpha$ denotes the observer angle with respect to the upward beam direction. For an observer angle $\alpha=90^\circ$, the antennas are located in the horizontal plane at the same height as the beam exit. Smaller viewing angles are obtained by elevating the experiments with respect to the height of the beam exit.

The electron beam sudden appearance signal was first found by the AirRadar experiment (50~MHz)~\cite{AirRadar}. The AirRadar experiment is performed in July 2012 to search for the in-air radar echo from the electron beam induced by the TA-ELS, using a roughly 5~ns bunch length. A wide-band (50~MHz-1.3~GHz) receiver system consisting of a log-periodic antenna was used to obtain the signal in the 50-66~MHz band. The transmitter and receiver are installed at a distance of 140~m from the beam exit at a horizontal viewing angle (90$^\circ$). 

The ARAcalTA experiment (230-430~MHz)~\cite{ARAcalTA} was performed in January 2015 to characterize the Askaryan effect as well as to confirm the ARA detector simulation and calibration~\cite{ARA}. The bunch length for this experiment was decreased to roughly 1.5~ns. The radio signal is collected by the same bi-cone antennas as those used for the ARA experiment. The antennas were located at a horizontal distance of 7.3~m from the beam exit, while the elevation of the antennas could be altered to obtain viewing angles between $45^\circ$-$90^\circ$.  

The IceRadar measurements (1.4-3~GHz)~\cite{IceRadar} were performed along with the ARAcalTA experiment in January 2015. This experiment was performed to investigate the radar detection technique to probe high-energy particle cascades in ice~\cite{IceRadar2}. The bunch length is the same as for the ARAcalTA experiment, roughly 1.5~ns. Data was taken using wifi-bow antennas, located at 2.83~m from the beam exit at a horizontal viewing angle (90$^\circ$). For this work, the low noise 2.1-2.3~GHz and 2.5-2.7~GHz bands are used. 

The Konan experiment (12.5~GHz) was performed in November 2014, and aimed to characterize Molecular Bremsstrahlung Radiation from the ionization plasma induced by the TA-ELS beam~\cite{Konan}. The bunch length for this experiment was similar to the 50~MHz experiment, roughly 5~ns. A feed-horn receiver was placed at 1.64 m from the beam exit with
a 35$^\circ$ viewing angle.

\section{Data and simulations}
\seclab{data-sim}
Besides the signal sought for, all four experiments detected a strong transient signal when the electron beam exits the accelerator. The data presented in this work is obtained using an experimental configuration without any (ice) target on top of the beam exit, directing the beam into the atmosphere. Furthermore, the radar transmitters were not operating for the presented data. The raw data observed by the different experiments are shown in~\figref{raw}. 

\begin{figure}[!ht]
\centerline{
\includegraphics[width=.45\textwidth, keepaspectratio]{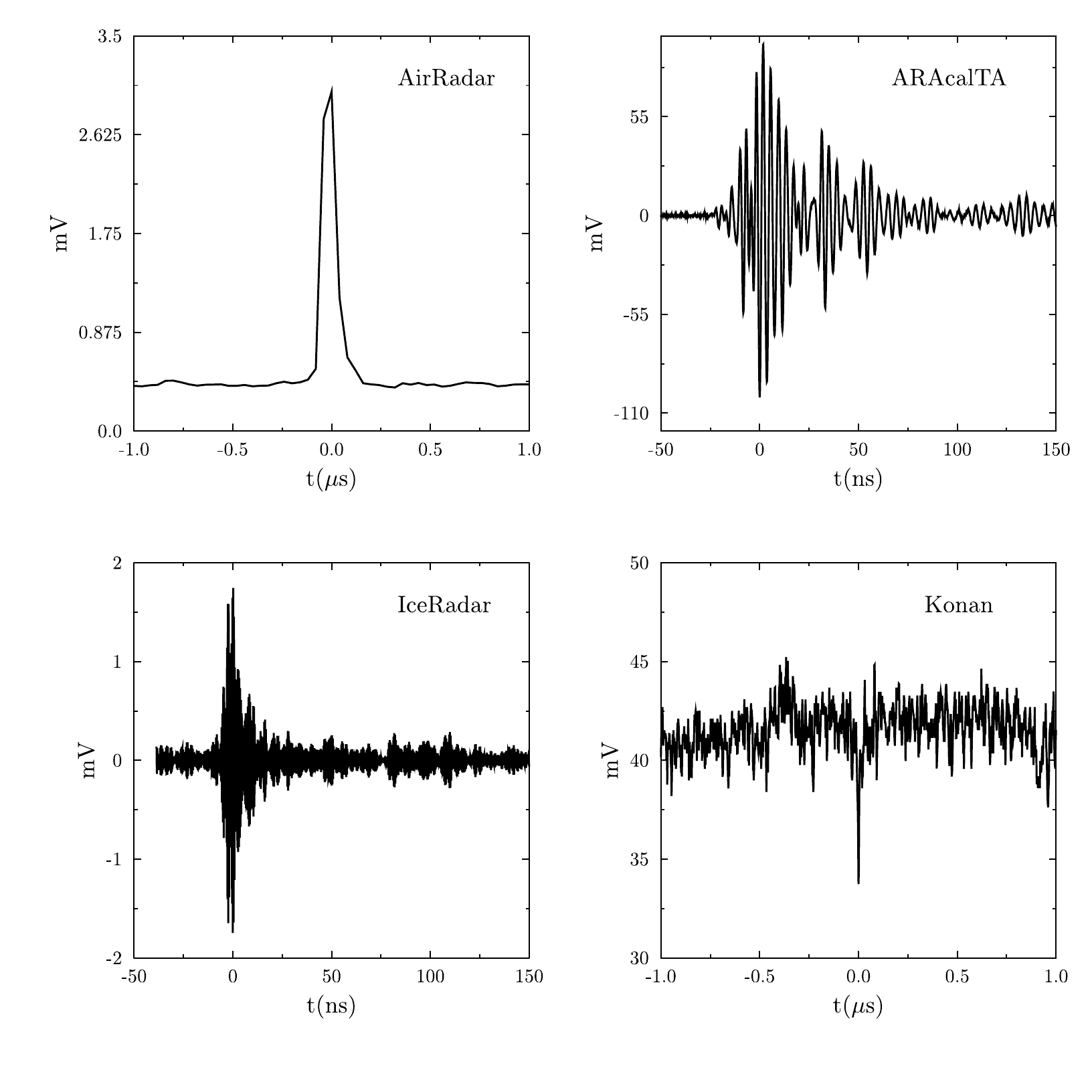}}
\caption{The raw data for the different experiments. AirRadar (50~MHz): top left~\cite{AirRadar}, ARAcalTA (230-430~MHz): top right~\cite{ARAcalTA}, IceRadar (2.1-2.3~GHz ; 2.5-2.7~GHz): bottom left~\cite{IceRadar}, Konan (12.5~GHz): bottom right~\cite{Konan}. The ARAcalTA and IceRadar data show the measured voltage for the radio frequency signals after filtering and amplification, while the AirRadar and Konan experiments measured the power integrated within their respective frequency bands. For the latter experiments, the readout values in units of Volt at the DAQ give a proxy for the signal power.}
\figlab{raw}
\end{figure}
\begin{table*}
\centering
\begin{tabular}{c|c|c|c}
\hline \hline
Setup & $E_{\mathrm{meas}}\mathrm{(J/m^2/Hz/pC^2)}$& $E_{\mathrm{sim}}\mathrm{(J/m^2/Hz/pC^2)}$ & $s$ \\
\hline 
50~MHz   & $1.00\;\pm\;0.01\;\mathrm{(stat)}\;^{+5.17} _{-0.56}\;\mathrm{(sys)}\;\times 10^{-24}$   & 3.16 $\pm$ 0.06 (sys) $\times$ $10^{-24}$ & $1.64 \pm 0.42$ \\
230-430~MHz  & 1.10 $\pm$ 0.09 (stat) $\pm$ 0.30 (sys) $\times$ $10^{-24}$&  1.34 $\pm$ 0.23 (sys) $\times$ $10^{-24}$&  $1.86\pm0.04$ \\
2.1-2.3 GHz   & 0.97 $\pm$ 0.03 (stat) $\pm$ 0.73 (sys) $\times$ $10^{-27}$ &  2.12 $\pm$ 0.36 (sys) $\times$ $10^{-27}$ &  $1.52\pm0.11$\\
2.5-2.7 GHz   & 3.02 $\pm$ 0.05 (stat) $\pm$ 2.27 (sys) $\times$ $10^{-27}$ &  4.59 $\pm$ 0.78 (sys) $\times$ $10^{-27}$ &  $2.02\pm0.07$\\
12.5~GHz  &  $ 8.40 \; \pm \; 4.42 \mathrm{(stat)} \; \pm\; 4.27\; \mathrm{(sys)}\; \times 10^{-29}$ & $>$ 2.37 $\times$ $10^{-32}$ &  $1.92\pm0.43$\\
\hline \hline
\end{tabular}
\caption{The observed energy density $E_{\mathrm{meas}}$ at the central frequency bands 50~MHz, 330~MHz, 2.2~GHz, 2.6~GHz, and 12.5~GHz, as well as the simulated energy density $E_{\mathrm{sim}}$, and the power index of the charge dependence $s$.}
\label{table1}
\end{table*}
\begin{figure}[!ht]
\centerline{
\includegraphics[width=.45\textwidth, keepaspectratio]{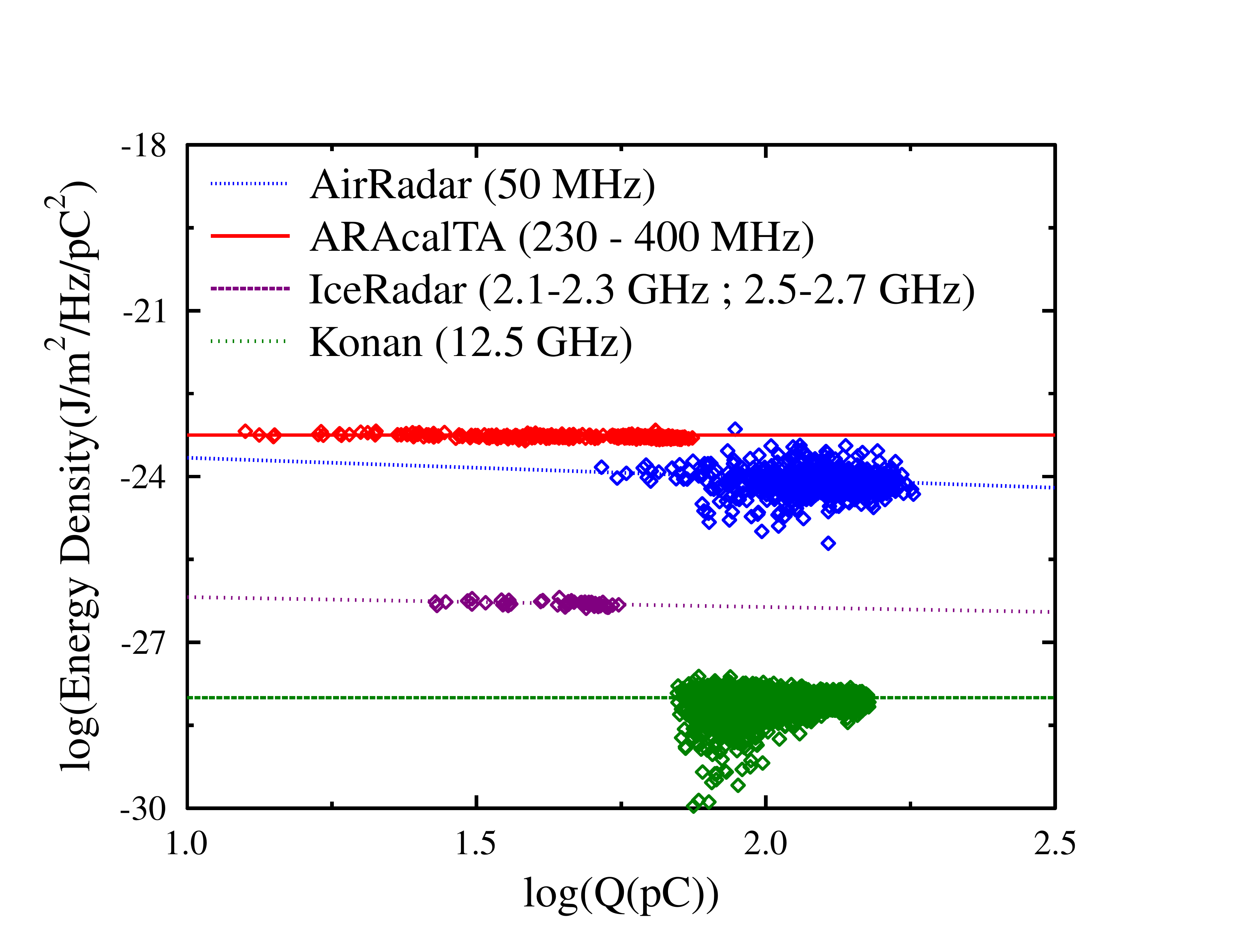}}
\caption{The obtained energy density normalized by the squared beam charge as function of beam charge for the AirRadar (blue points), ARAcalTA (red points), IceRadar (purple points) and Konan (green points) experiments. For the given normalization, no charge dependence is expected for fully coherent emission ($s=2$). The obtained coherence levels for the different experiments are given in Table~1.}
\figlab{coherence}
\end{figure}
The energy density normalized by the beam charge at the central frequency for each experiment is given in Table~\ref{table1}. Since the different measurements were taken at different positions with respect to the emission point, a distance correction is made to compare the obtained results. To obtain its expected value at 1~m from the beam exit we assume the emission energy to follow an inverse square distance dependence, which is valid for emission in the far-field.

For the ARAcalTA measurement we consider the data obtained under a horizontal viewing angle. The main systematic error for the ARAcalTA, and IceRadar experiments is due to the uncertainty in the total system gain. The directional AirRadar antenna was not directed (horizontally) at the beam appearance point at the top of the cover box, but pointing 10 degrees above the horizontal~\cite{AirRadar}. As such, the beam sudden appearance was observed on the edge of the antenna gain pattern leading to a large asymmetry in the systematic error. The main error for the Konan experiment is given by the accelerator noise, which correlates with the obtained signal.

The obtained data in combination with the beam charge obtained from the WCM allows us to check the level of coherence for the different experiments. The charge dependence of the radio energy density is fitted to follow a power-law $E=AQ^{s}$, where the power index $s$ gives the coherence level. A fully incoherent signal would give a linear scaling of the signal with charge ($s=1$), while a quadratic scaling ($s=2$), is expected for full coherence. The obtained values for the power index are given in the fourth column of Table~\ref{table1}. The fit, as well as the obtained data is shown in~\figref{coherence} where we plot the energy density as function of the beam charge. All experiments show a high level of coherence.

The predicted signal for the coherent electron beam sudden appearance is derived from the Li\'enard-Wiechert potentials of classical electrodynamics by imposing the potential to vanish while the beam propagates inside the accelerator. This formalism closely follows the modeling of coherent transition radiation and sudden appearance emission outlined in Sec.~2, described both macroscopically in~\cite{CosmicRayRadio}, and microscopically in~\cite{Motloch}. 
\begin{figure}[!ht]
\centerline{
\includegraphics[width=.53\textwidth, keepaspectratio]{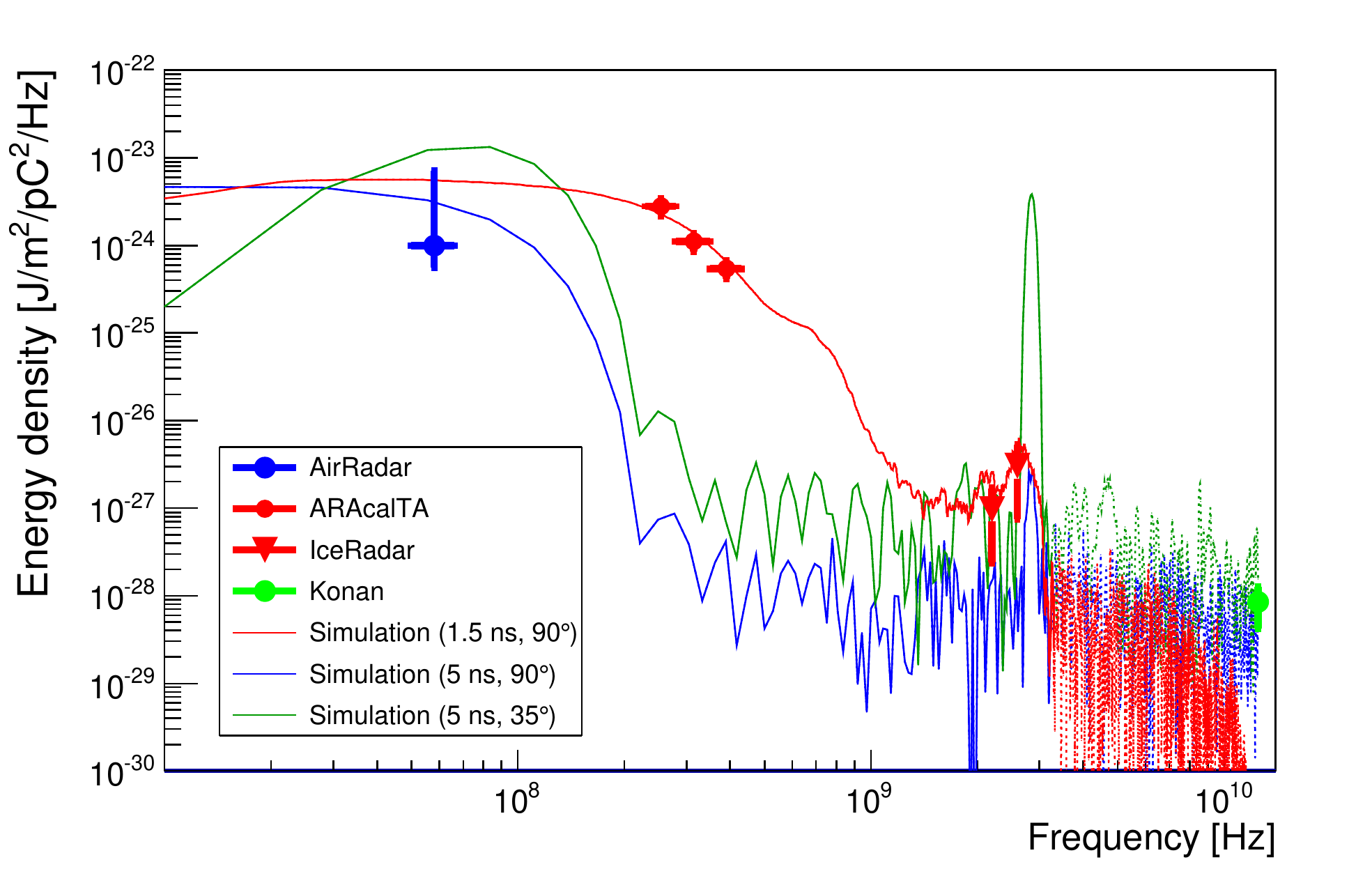}}
\caption{Observed data points and simulated energy densities for the different experiments: AirRadar (blue dot, blue line) ; ARAcalTA (red dots, red line) ; IceRadar (red triangles, red line) ; Konan (green dot, green line). At low frequencies (full lines) coherence is obtained over the longitudinal particle structure. At the highest frequencies the lateral particle distribution becomes the leading length scale (dashed lines).}
\figlab{eden}
\end{figure}

Simulations are performed using the individual particle trajectories obtained from GEANT4 simulations~\cite{GEANT4}, after which the radio emission from the individual tracks is obtained using the ZHS algorithm~\cite{ZHS}. These simulations have been cross-checked with the macroscopic approach given in~\cite{CosmicRayRadio}.

The third column of Table~\ref{table1} gives the expected energy density obtained from our simulations. The predicted energy density spectrum for each experiment is also shown in~\figref{eden}. The spectrum directly reflects the spatial distributions of the particle beam. The main coherence arises from the total beam length. The loss of coherence leads to a cut-off around 200-500~MHz in the simulation for the ARAcalTA and IceRadar experiments, using a 1.5~ns bunch length and an observation angle of 90$^\circ$  (red line). A cut-off around 20-50~MHz is obtained using a 5~ns bunch length and observation angles of 90$^\circ$ for the AirRadar experiment (blue line), or 35$^\circ$ for the Konan experiment (green line). The coherence from the 350~ps sub-bunch structure shows up around 2.856~GHz. At the highest frequencies ($>3$~GHz), coherence over the beam bunch length is lost and the leading length scale is given by the lateral particle distribution which is found to be a few cm.

The errors on the simulated values are due to the uncertainty in the particle distributions within the electron beam. For the  AirRadar, ARAcalTA and IceRadar experiments the main simulation uncertainty is given by the uncertainty in the beam bunch length. The beam bunch length is determined from the Faraday cup measurements, providing an average error of $15\%$ for the 1.5~ns beam bunch configuration, and $2\%$ for the 5~ns bunch length, giving stable simulation results below 3~GHz indicated by the full lines in Fig.~4. For the Konan experiment, coherence is determined at cm scales, and the lateral particle distribution becomes the leading uncertainty. As the sub-structures within the lateral particle distribution are unknown, a lower-limit on the predicted field is given in Table 1. Since we are only able to give a lower-limit on the predictions at the highest frequencies ($>3$~GHz), in this frequency range the results shown by the dashed lines in Fig.~4 are fitted to the Konan data by modifying the fine structure of the lateral particle distribution.



The experimental results are also shown in~\figref{eden}. The results are consistent with the predicted signals over a wide range of frequencies, ranging over many orders of magnitude. The emission follows the expected behavior given the beam charge profile for each experiment.

The angular dependence of the sudden appearance signal is probed using the ARAcalTA data which was taken for 9 different heights corresponding to emission angles from 45$^{\circ}$ to 90$^{\circ}$. The measured radio energy as function of the emission angle is shown in~\figref{fig:angdist}, showing good agreement with simulations. 

From the results presented in this section, it follows that our measurements are in good agreement with the electron beam sudden appearance hypothesis, for which the potentials are effectively zero while the electron beam is propagating inside the container.
\begin{figure}[!ht]
  \centering
  \hspace*{-3ex}  	
  \includegraphics[width=0.9\linewidth]{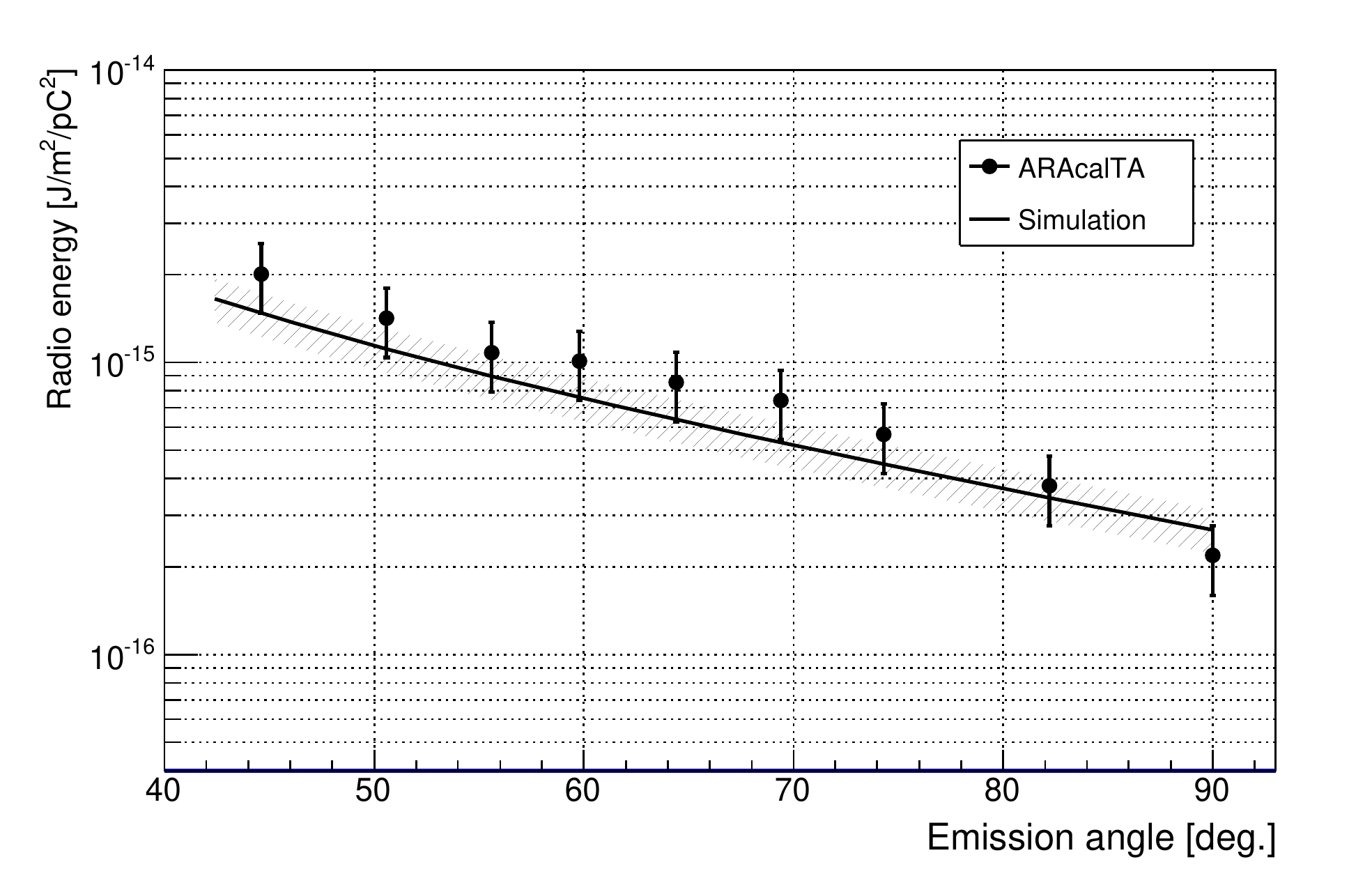}	 
  \caption{Angular dependence of the sudden appearance radio energy for the ARAcalTA experiment. The shaded area indicates the simulation with its uncertainty (1$\sigma$) given by the uncertainty in the beam bunch length.}  
  \figlab{fig:angdist}
\end{figure}

\section{Summary}
We investigated the electron beam sudden appearance signal at the TA-ELS facility using data of four different experiments ranging from 50~MHz up to 12.5~GHz. All experiments measured a clear transient signal when electron beam exits the accelerator container. The observed signals show a high level of coherence, reflecting the details of the particle distributions within the electron beam. We show that the expected signal can be understood as a special case of coherent transition radiation. This is confirmed by the agreement between simulation and data over the full frequency range. The angular dependence of the signal is obtained by the ARAcalTA experiment, showing a good agreement with simulations.

A very interesting conclusion which can be drawn from the obtained results is that the electron beam does not have to cross a physical boundary layer between two different media to give strong, transition radiation like, emission. It merely has to move from one environment (the accelerator container) to another (free space) from the observer point of view. This effect can be understood by considering the potential which is observed while the beam propagates inside the container. In this situation, the beam emission effectively vanishes for an outside radio detection set-up. It follows that at the container exit, the electron beam suddenly appears for the observer, which leads to a strong shock in the potential, and hence the observed field. 

In this article, we for the first time show that simulations predicting the coherent transition radiation signal are in good agreement with data over a wide range of radio frequencies from 50~MHz up to 12.5~GHz. These simulations show that the coherent transition radiation signal should be observable from cosmic ray air showers penetrating the Antarctic ice, using already installed radio detectors~\cite{CosmicRayRadio}. Even though a first search was performed recently~\cite{Arianna_nelles}, up to now no significant detection of such a process in nature has been claimed. The reverse process focusing on coherent transition radiation from neutrino induced cascades moving from dense media to air was considered in~\cite{Motloch}. The hypothesis that the upward going cosmic-ray like events detected by the ANITA detector~\cite{ANITA1,ANITA2} are due to coherent transition radiation is investigated in~\cite{Motloch2}. It is shown that even though the properties of the signal detected by ANITA can be explained by coherent transition radiation, the neutrino flux needed to produce this signal is in tension with the current experimental upper limits. Besides these detectors, a more dedicated surface experiment is currently under construction to detect the coherent transition radiation from a cosmic-ray air shower hitting Earth's surface, focusing on the so-called air-shower sudden death signal~\cite{extasis}.

\section{Acknowledgments}
The authors would like to thank the TA collaboration for the use of the TA-ELS facility and continuous supports to realize the experiments. We thank KEK accelerator scientists in the electron/positron injector
linac group for valuable advices in operating the ELS. The authors also would like to thank Anne Zilles for her kindness to share her source code to implement the ZHS method for the simulations. Furthermore, the authors would like to thank Olaf Scholten for the very helpful and insightful discussions. 
This work was supported by the Flemish Foundation for Scientific Research FWO (FWO-12L3715N - K. D. de Vries; G.0917.09. - N. van Eijndhoven);
by the Japan Society for the Promotion of Science (JSPS) (JP23654078 and JP25400296 - D. Ikeda; JP25220706 - A. Ishihara, K. Mase and S. Yoshida;
15K13489 - T. Yamamoto); by the joint research program of the Institute for Cosmic Ray Research (ICRR), The University of Tokyo.

\end{document}